\documentclass[journal]{IEEEtran}
\usepackage{cite}
\usepackage{epsfig}
\usepackage[cmex10]{amsmath}
\usepackage[tight,footnotesize]{subfigure}
\def\e{\begin{equation}}
\def\f{\end{equation}}
\usepackage{url}
\usepackage{color}

\begin{document}
\title{Experimental Characterization of a Broadband Transmission-Line Cloak in Free Space}

\author{Pekka Alitalo, Ali E. Culhaoglu, Andrey V. Osipov, Stefan
Thurner, Erich Kemptner, and Sergei~A.~Tretyakov% <-this % stops a space
\thanks{P. Alitalo and S.~A.~Tretyakov are with the Department of Radio Science and Engineering/SMARAD Centre of Excellence, Aalto University School of Electrical Engineering, P.O. Box 13000 FI-00076 Aalto, Finland. Email: pekka.alitalo@aalto.fi}%
\thanks{A. E. Culhaoglu, A. V. Osipov, S. Thurner, and E. Kemptner
are with the Microwaves and Radar Institute, German Aerospace
Center (DLR), 82234 Wessling, Germany.}
\thanks{This work has been partly funded by the Academy of Finland and Nokia through
the centre-of-excellence program. The work of P. Alitalo has been
supported by the Academy of Finland through post-doctoral project
funding.}
\thanks{P. Alitalo acknowledges the work of Mr. Eino
Kahra in helping in the manufacturing of the cloak.}}

\maketitle


%\begin{IEEEkeywords}
%Scattering, scattering cross section.
%\end{IEEEkeywords}

%\ifCLASSOPTIONpeerreview
%\begin{center} \bfseries EDICS Category: 3-BBND \end{center}
%\fi
\IEEEpeerreviewmaketitle

\begin{abstract}
%A finite-sized, cylindrical transmission-line cloak operating in
%the X-band is characterized with bistatic free-space measurements.
%The cloak design and dimensions are optimized numerically and the
%cloaking effect is studied with numerical full-wave simulations.
%The cloaking effect is also verified with extensive bistatic
%measurements from which the reduction of the total scattering
%width, enabled by the cloak, can be clearly observed. The
%numerical and experimental results are compared resulting in good
%agreement with each other.

The cloaking efficiency of a finite-size cylindrical
transmission-line cloak operating in the X-band is verified with
bistatic free space measurements. The cloak is designed and
optimized with numerical full-wave simulations. The reduction of
the total scattering width of a metal object, enabled by the
cloak, is clearly observed from the bistatic free space
measurements. The numerical and experimental results are compared
resulting in good agreement with each other.

\end{abstract}

\maketitle

\section{Introduction}

The transmission-line cloak concept has been recently
introduced~\cite{TAP2008,ProcIEEE} as an alternative to the
transformation-optics~\cite{TO,TOreview} and scattering
cancellation approaches~\cite{SC,SCreview}. Instead of utilizing
anisotropic (and often resonant) metamaterials~\cite{TOreview} or
plasmonic materials~\cite{SCreview}, the transmission-line cloak
enables the electromagnetic wave to smoothly travel through the
cloaked object inside a volumetric network of transmission lines,
resulting in a simple and cheap way to obtain broadband cloaking
of objects with \textit{selected} geometries. It should be
emphasized that the transmission-line cloak can only ``hide''
objects that fit inside the volumetric network~\cite{ProcIEEE}.

A clear distinction should be made between three-dimensional
cloaks that can hide an object in free space and the so-called
ground-plane cloaks that can be used to hide an object above a
boundary~\cite{GPcloak1,GPcloak2}. The ground-plane cloaks are
eminently simpler to realize since the complexity of the material
parameters is not so demanding as in cloaks that should hide a
finite object in free space.

In this work we study a finite-size, three-dimensional
transmission-line cloak that can hide a three-dimensional metallic
object from electromagnetic waves in free space. The basic cloak
geometry is known from previous results~\cite{ProcIEEE} and the
dimensions of the cloak are here optimized for operation in the
X-band (8~GHz -- 12~GHz). The previous realizations of the
cylindrical transmission-line cloak utilized a coupling layer made
of widening metal strips to couple the electromagnetic waves
between the free space and the volumetric transmission-line
network. To simplify the cloak design and to enable more accurate
and faster assembly of the structure, we have very recently
introduced an alternative way of realizing the coupling layer by
using solid conical metal discs~\cite{EUCAP2011} instead of
strips. This makes both the design and manufacturing of the cloak
simpler since the structure is more rigid and the coupling layer
does not anymore bring a strong frequency dependence to the cloak
operation~\cite{EUCAP2011}.

In this paper we introduce the new cloak design and present
numerical results for the ratio of the total scattering widths of
cloaked and uncloaked objects. The cloak is manufactured and its
operation is experimentally verified with an X-band bistatic
measurement setup consisting of two antennas and a vector network
analyzer (VNA). With this setup we can obtain experimentally the
ratio of the total scattering widths that can then be compared to
the numerical results. It is shown that both the numerical and
experimental results demonstrate a broadband cloaking effect. Free
space bistatic measurements conducted with a finite-size cloak are
an important step toward a full characterization of practical
electromagnetic cloaks, as the previously published experimental
results considering total scattering from various cloaks have been
conducted with parallel-plate waveguide setups~\cite{SCexp,TOexp}.

\section{Cloak Geometry}

\begin{figure} [t!]
\begin{center}
\subfigure[]{\epsfig{file=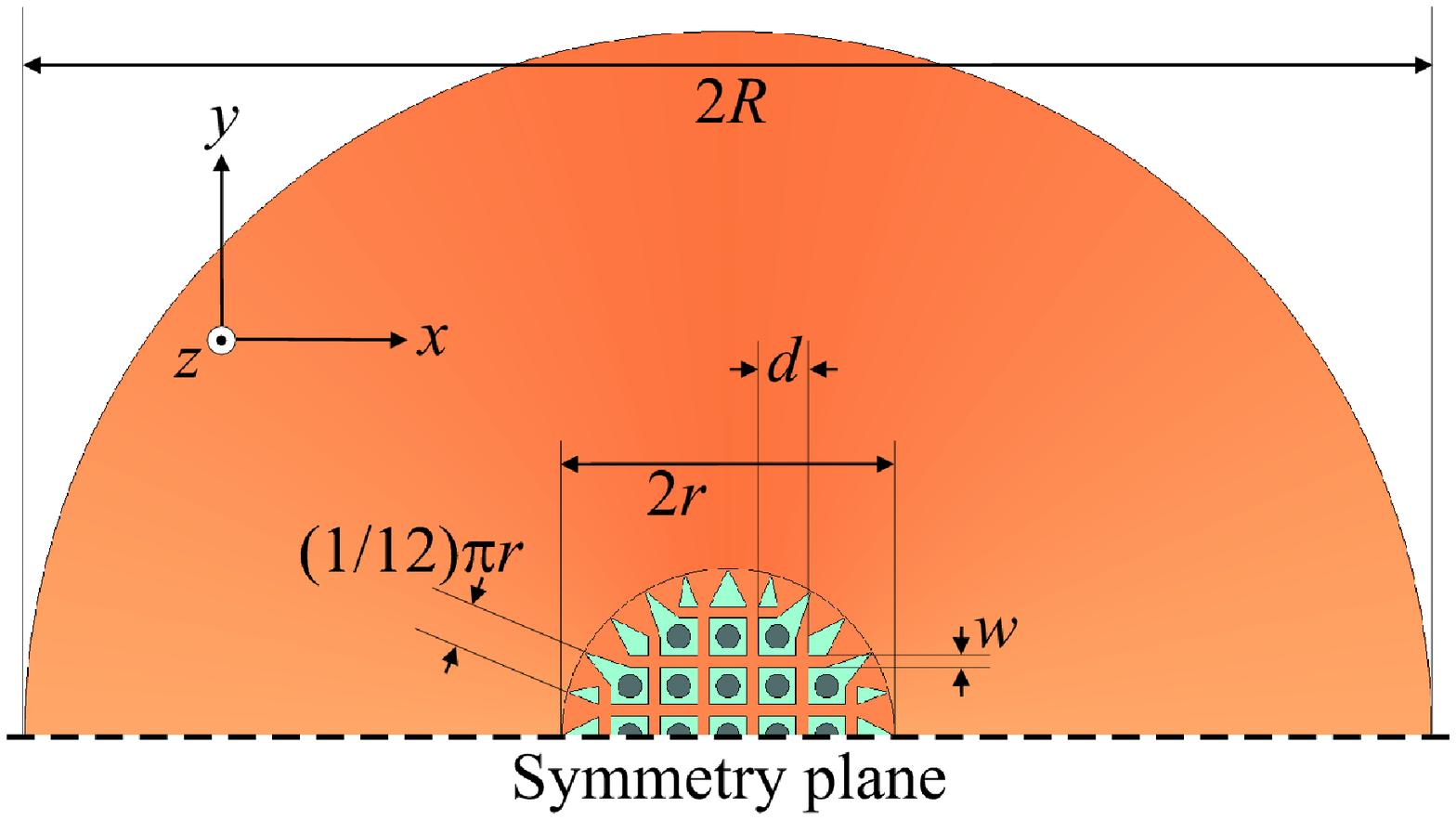, width=0.4\textwidth}}
\subfigure[]{\epsfig{file=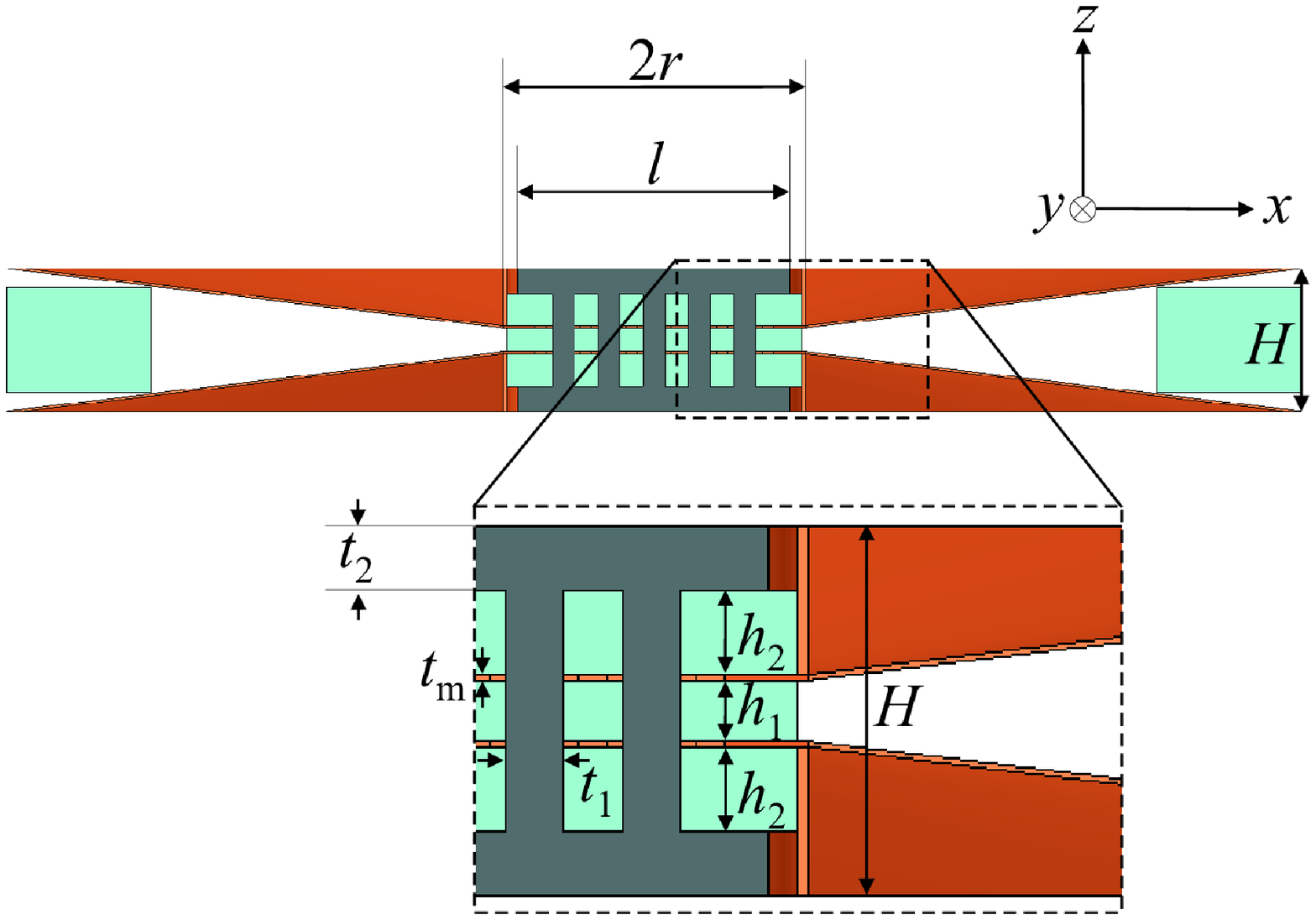, width=0.4\textwidth}}
\subfigure[]{\epsfig{file=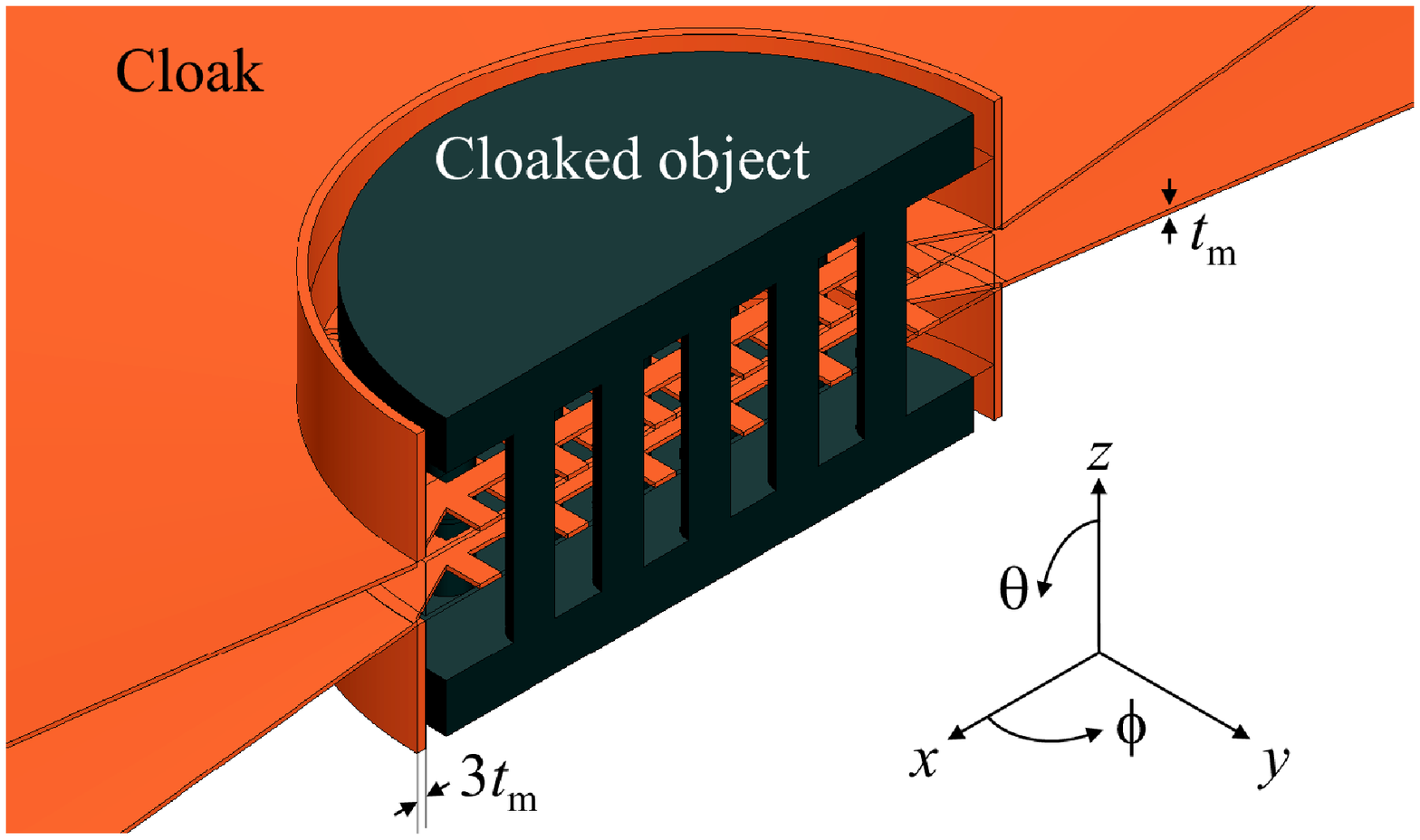, width=0.4\textwidth}}
\caption{Geometry and dimensions of the transmission-line cloak.
The cloak is periodic along the $z$-axis. (a)~$xy$-plane view.
(b)~$xz$-plane view. (c)~One half of a single period of the cloak,
cut along the $xz$ symmetry plane (dielectric not shown for
clarity).}\label{geometry}
\end{center}
\end{figure}

The present cloak geometry is a modification of the previously
studied cylindrical transmission-line cloak~\cite{ProcIEEE}. The
transition or coupling layer that couples the electromagnetic
waves between a set of layered two-dimensional transmission-line
networks and the surrounding free space, is composed of metal
cones, as illustrated in Fig.~\ref{geometry}. This coupling layer
geometry has been studied numerically~\cite{EUCAP2011} and it was
shown that, unlike in the previous designs utilizing gradually
widening strip transmission lines~\cite{ProcIEEE}, the outer
diameter of the cone (i.e., the thickness of the transition layer)
does not strongly affect the frequency of optimal cloaking. The
coupling and therefore the cloaking performance are improved as
the thickness of this layer increases~\cite{EUCAP2011}. For
practical purposes we limit the thickness of these layers after
obtaining a good level of cloaking so that the cloak does not
become impractically large compared to the cloaked object.

The optimization of the cloak dimensions was carried out
essentially by first directly scaling the dimensions of the
previous structure~\cite{EUCAP2011}, operating at frequencies
around 3~GHz, so that the new operating frequency band occurs in
X-band with the center of the cloaking band at 10~GHz. Then,
taking into account material restrictions (the dielectric material
used for insulation between metal parts was available only in
certain specific thicknesses), some design parameters were fixed
to appropriate values regarding manufacturing and assembly while
the remaining few key parameters (e.g. the height of a single
cloak layer $H$) were numerically optimized with ANSYS
HFSS~\cite{HFSS}.

As illustrated in Fig.~\ref{geometry}, the cloak consists of thin
sheets of metal and a dielectric insulator (heights $h_1$ and
$h_2$) that is used to separate the successive layers of
transmission-line strips from each other and from the cloaked
metal object. As shown in Fig.~\ref{geometry}, the cloaked object
is an array of long metal rods connected periodically to a set of
thin metal discs. It should be emphasized that in principle there
is no restriction on the geometry of the cloaked object (nor, for
example, the number of rods), as long as it fits inside the
volumetric transmission-line network~\cite{ProcIEEE}. As it is
favourable to have the wavenumber of the transmission lines as
close as possible to the wavenumber in free space~\cite{ProcIEEE},
the dielectric is chosen to have low permittivity and low losses.
Here we use Rohacell$^{\textregistered}$ 51 HF~\cite{Rohacell}
with a relative permittivity of $\varepsilon_{\rm r}=1.07$ and
loss tangent of 0.003.

\begin{table}[t!]
  \centering
  \caption{Dimensions of the transmission-line cloak.}
  \begin{tabular}{c|c p{2.0cm}}
    \hline

    $d$ & 2.1 mm \\\hline
    $w$ & 0.5 mm \\\hline
    $h_1$ & 1.1 mm \\\hline
    $h_2$ & 1.5 mm \\\hline
    $H$ & 6.6 mm \\\hline
    $r$ & 7 mm \\\hline
    $R$ & 30 mm \\\hline
    $l$ & 12.6 mm \\\hline
    $t_{\rm m}$ & 0.1 mm \\\hline
    $t_{\rm 1}$ & 1 mm \\\hline
    $t_{\rm 2}$ & 1.15 mm \\\hline
  \end{tabular}
  %\vspace{-1cm}
\end{table}

\begin{figure} [t!]
\begin{center}
\subfigure[]{\epsfig{file=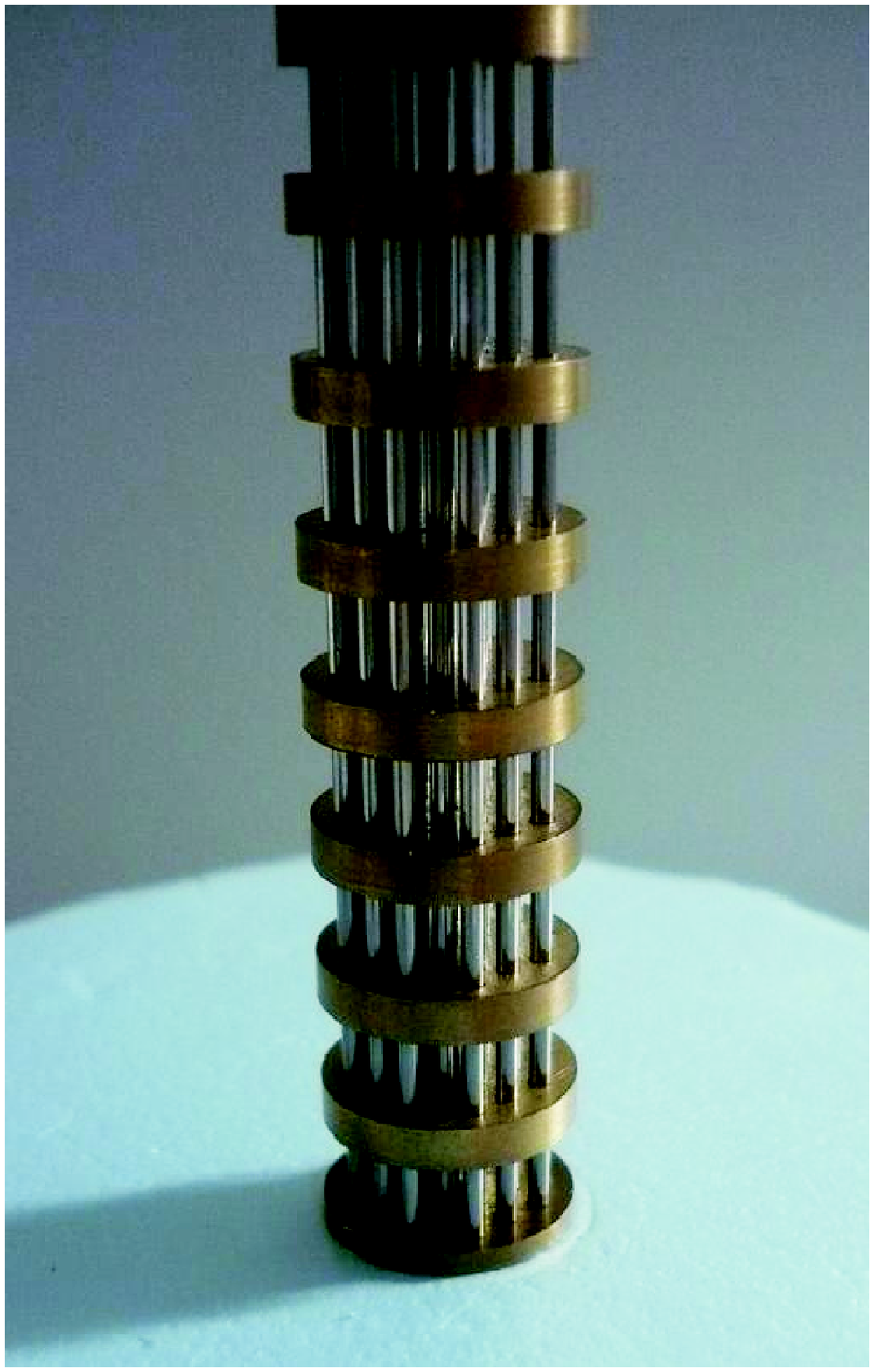, width=0.15\textwidth}}
\subfigure[]{\epsfig{file=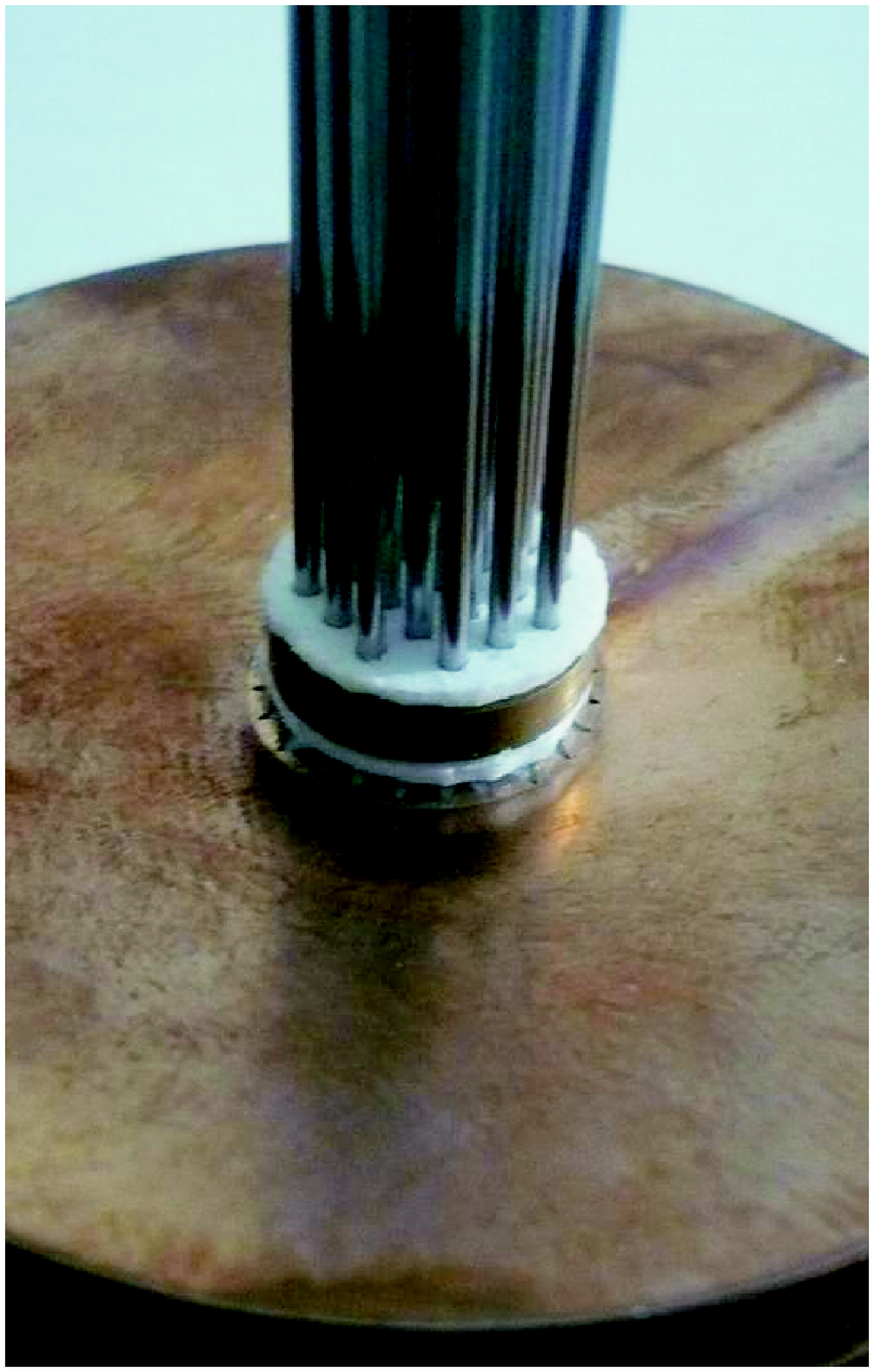, width=0.15\textwidth}}
\subfigure[]{\epsfig{file=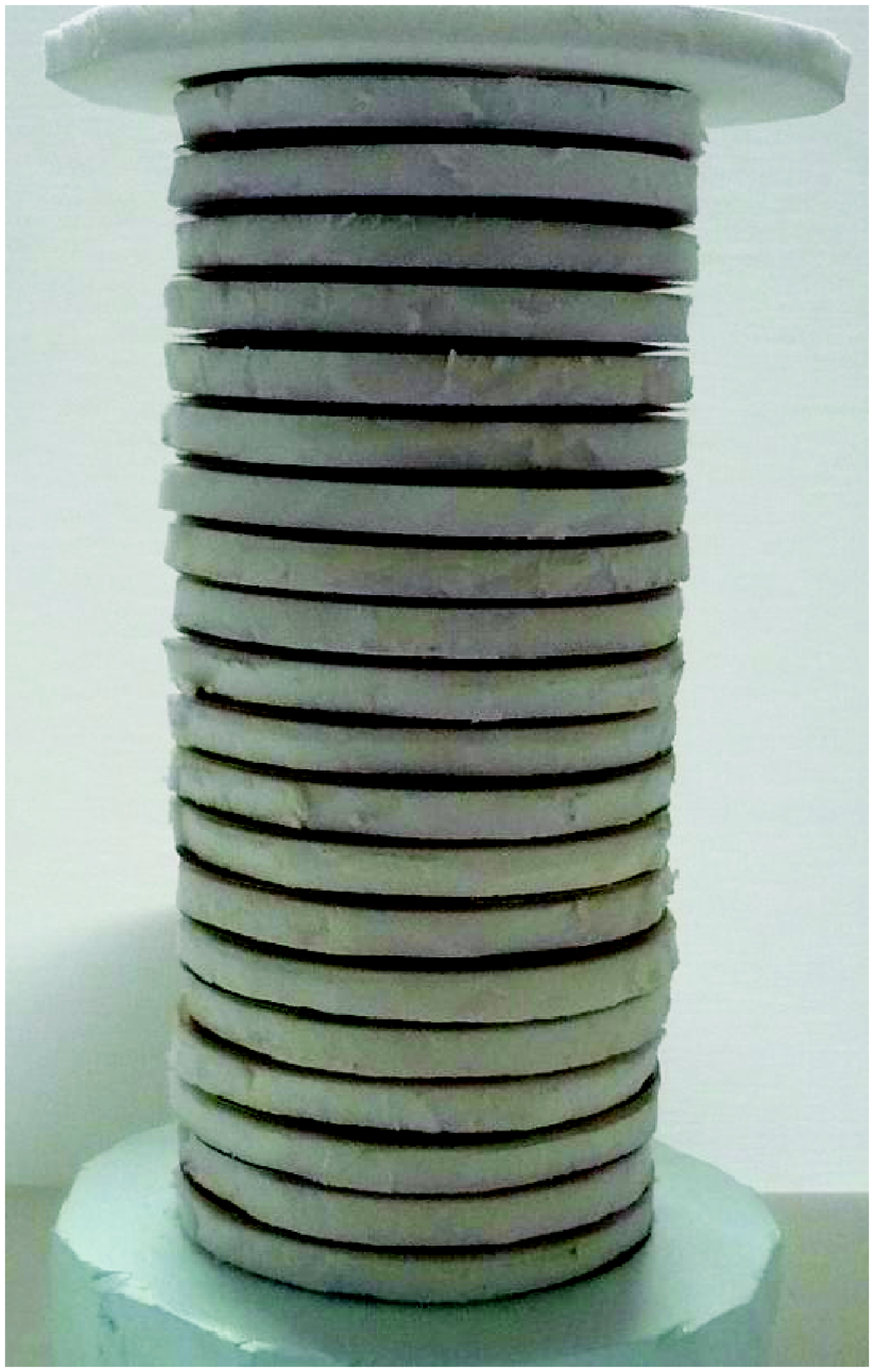, width=0.15\textwidth}}
\caption{Photographs of (a)~uncloaked object (only partly shown),
(b)~cloaked object in assembly, (c)~measured cloaked object.}
\label{assembly}
\end{center}
\end{figure}

For practical reasons we do not employ solid metal cones in the
structure, but instead thin metal sheets that are bent to the
conical shape. To obtain the same response as with solid cones, we
place a hollow metal cylinder (with thickness $3t_{\rm m}$ and
outer radius $r$, see Fig.~\ref{geometry}b,c) around the cloaked
object so that the inside volume of the cone is impenetrable to
electromagnetic waves. Photographs of the structure are shown in
Fig~\ref{assembly}. The final, finite-size cloak consists of 20
layers of the structure of Fig.~\ref{geometry}, resulting in the
total height of $20\times 6.6$~mm~$=132$~mm for both the cloaked
and uncloaked object. This height is much larger than the
half-power width of the antenna pattern at the location of the
object in the measurement setup (see the next section), which
means that we can expect negligible reflections from the top and
bottom of the cloak structure and we can assume that the structure
scatters effectively in the $xy$-plane only.

Since the metal cones are made of thin sheets, the final cloak
structure is made mechanically more stable by adding cylindrical
dielectric supports to the ends of the openings of the metal cones
(illustrated in Fig.~\ref{geometry}b and Fig.~\ref{assembly}c).
These dielectric supports are also made of the same
Rohacell$^{\textregistered}$ material as above and have the height
equal to 5~mm and the inner and outer radii of 23.5~mm and 30~mm,
respectively. Although the effect of these support cylinders is
expected to be very minor due to the low permittivity of the
material, these cylinders are taken into account in the numerical
simulations.

\section{Measurement Setup and Analysis}

\begin{figure} [t!]
\begin{center}
\subfigure[]{\epsfig{file=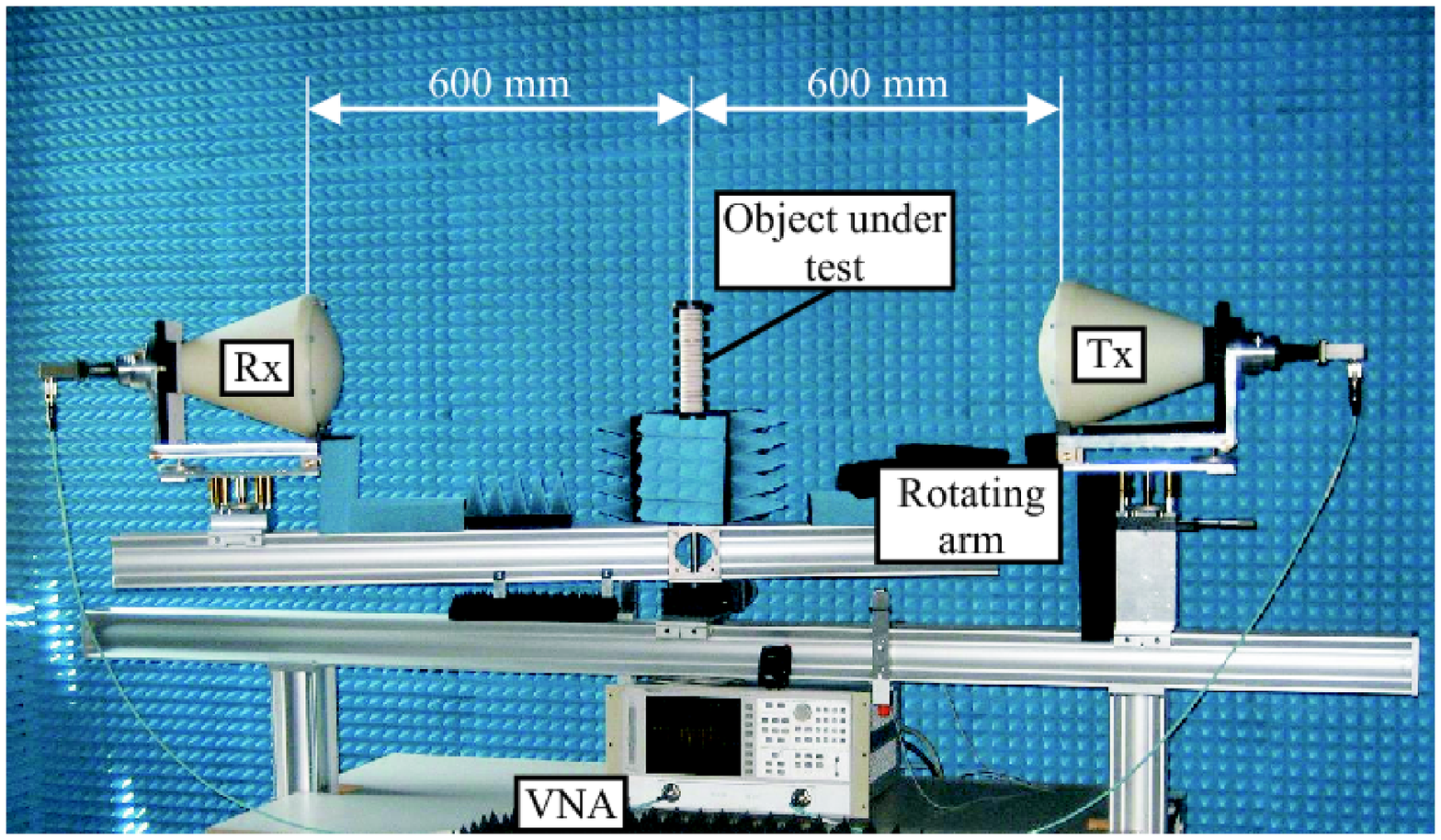, width=0.49\textwidth}}
\subfigure[]{\epsfig{file=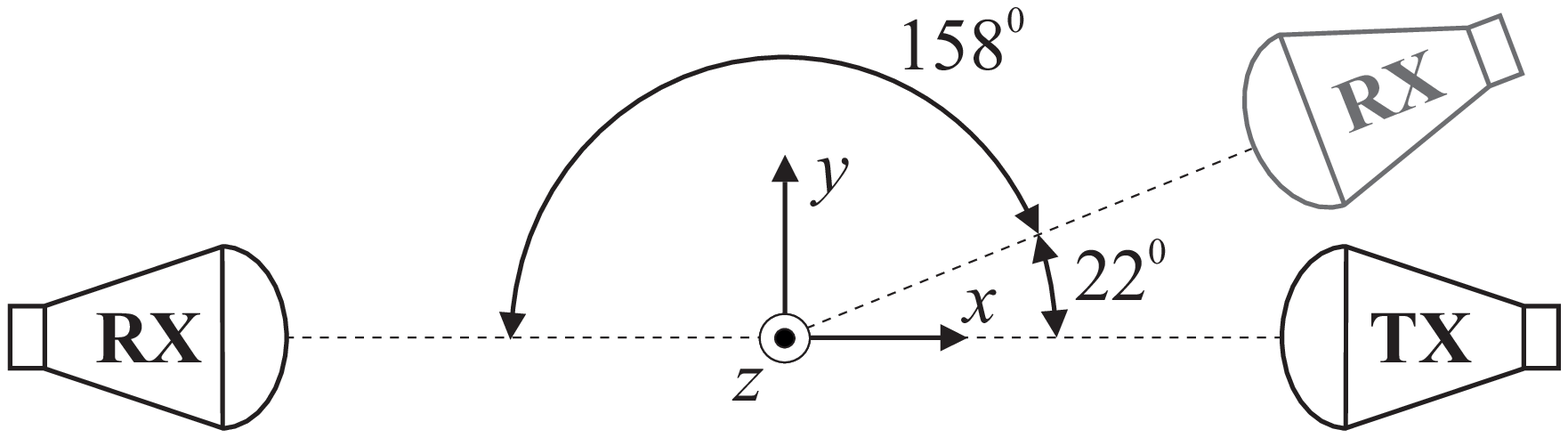, width=0.4\textwidth}}
\caption{Illustration of the measurement setup at the German
Aerospace Center (DLR). (a)~Photograph of the setup. (b)~The
transmitting antenna (Tx) is stationary whereas the receiving
antenna (Rx) can be rotated around the coordinate origin in an
angular range of $158^{\circ}$.} \label{setup}
\end{center}
\end{figure}

The measurement setup consists of two X-band antennas equipped
with dielectric lenses that focus the electromagnetic energy on a
focal spot (half-power width approximately 45~mm) between the
antennas. The object under test is placed in this focus as shown
in Fig.~\ref{setup}. The transmitting antenna (Tx) is stationary,
whereas the receiving antenna (Rx) is placed on a rotating arm
that can be moved around the measured object. %Due to spatial limitations in the anechoic chamber around the
%setup, the full $360^{\circ}$ angular range cannot be measured,
%but this is not a problem since the cloaked and uncloaked objects
%are symmetric along the $xz$-plane, so it would be enough to
%measure only the angular range from $0^{\circ}$ to $180^{\circ}$.
The measured objects are symmetric with respect to the $xz$-plane,
so that a bistatic measurement over the angular range from
$0^{\circ}$ to $180^{\circ}$ is sufficient. However, due to large
antenna dimensions, we are restricted to the angular range from
$22^{\circ}$ to $180^{\circ}$, as illustrated in
Fig.~\ref{setup}b. The measurements are conducted with an angular
step of $0.5^{\circ}$ and for each angle, the complex transmission
coefficient ($S_{\rm 21}$) between the antennas is measured with a
vector network analyzer (Agilent HP 8719D) in the frequency range
8.2~GHz -- 12.4~GHz.

To characterize the total scattering width of an object, we need
to know the angle-dependent scattered field $E_{\rm sca}(f,\phi)$.
This can be obtained from the measured transmission results by
subtracting the free space transmitted fields (transmission
between the antennas without any scattering object in between):

\begin{equation}
E_{\rm sca}(f,\phi) = S_{\rm 21,O}(f,\phi) - S_{\rm
21,FS}(f,\phi),
\end{equation} where $S_{\rm 21}$ is the transmission coefficient between
the Tx and Rx antennas, and ``O'' and ``FS'' refer to ``object''
and ``free space'', respectively. The normalized total scattering
width $\sigma_{\rm W,norm}$, i.e., the total scattering width of
the cloaked object normalized by the total scattering width of the
uncloaked object, can be obtained by integrating the scattered
field intensities:

\begin{equation}
\sigma_{\rm W,norm}(f) = \frac{\int{|E_{\rm
sca,cloaked}(f,\phi)|^2}d\phi}{\int{|E_{\rm
sca,uncloaked}(f,\phi)|^2}d\phi}.
\end{equation}

In addition to the bistatic measurements used in (1), one can also
make a monostatic measurement with only one antenna in order to
obtain the scattered field at $\phi=0^{\circ}$. This is done
simply by measuring the reflection coefficient ($S_{\rm 11}$) of
the Tx antenna and equating the result to the scattered field:

\begin{equation}
E_{\rm sca}(f,\phi=0^{\circ}) = S_{\rm 11,O}(f).
\end{equation} This quantity is not used in the integration of
(2) since it provides only one angle point. It is used to
illustrate the small level of scattered field close to the
backscattering direction.

%verify the angular dependency of scattered field by comparing the
%measured result to the numerical one.

\section{Numerical and Experimental Results}

\begin{figure} [t!]
\begin{center}
{\epsfig{file=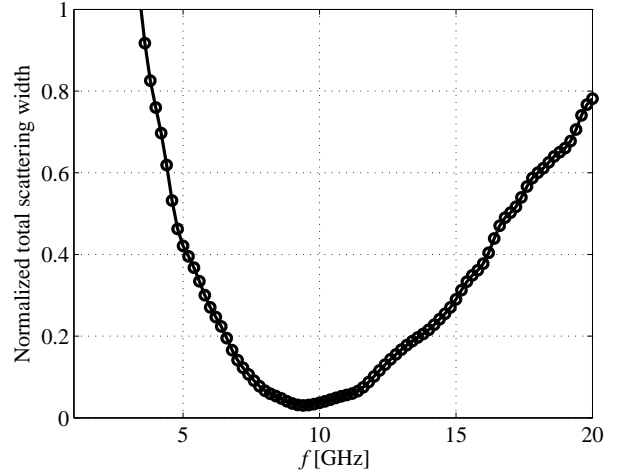, width=0.49\textwidth}} \caption{Numerical
result for the normalized total scattering width.} \label{TSW}
\end{center}
\end{figure}

The numerical analysis of the studied structures is carried out
with the commercial full-wave simulation software ANSYS
HFSS~\cite{HFSS}. The uncloaked and cloaked structures (see
Figs.~\ref{geometry} and~\ref{assembly}) are numerically modelled
as being infinitely periodic in the $z$-direction, by using
symmetry boundaries. The uncloaked and cloaked objects are
illuminated with a plane wave travelling in the $-x$-direction and
having the electric field parallel to the $z$-axis. From the
simulated fields one can obtain the scattered fields $E_{\rm sca}$
(1) and the resulting normalized total scattering width (2) as a
function of the frequency. The numerical result of (2) with
integrating from $0^{\circ}$ to $180^{\circ}$ is presented in
Fig.~\ref{TSW}, showing that the cloak effectively reduces the
total scattering width in a wide frequency band. In the band from
7.5~GHz to 11.9~GHz the total scattering width of the cloaked
object is less than 10~\% of the total scattering width of the
uncloaked object.

\begin{figure} [t!]
\begin{center}
{\epsfig{file=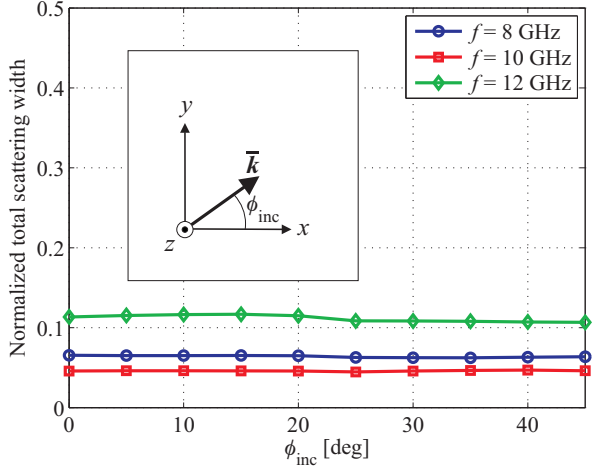, width=0.49\textwidth}}
\caption{Numerical results for the normalized total scattering
width with plane wave incident from various angles.}
\label{TSW_obl}
\end{center}
\end{figure}

The isotropy of the cloaking effect can be studied by varying the
incidence angle of the incoming plane wave. In this case we fix
the frequency and study the normalized total scattering width as
function of the incidence angle $\phi_{\rm inc}$, see
Fig.~\ref{TSW_obl}. Due to the symmetry of the structure, it is
enough to study an angular range of $45^{\circ}$, see
Fig.~\ref{geometry}a. Figure~\ref{TSW_obl} demonstrates that the
cloak behavior practically does not change with varying incidence
angle. It should be noted that only TE incidence with the electric
field parallel to the $z$-axis is considered here. The cloaking
effect can deteriorate for other polarizations.
%It should be noted though that here we are always keeping the TE
%incidence, i.e., the electric field is always parallel to the
%$z$-axis. The cloaking effect can of course be expected to
%deteriorate if this condition is not satisfied.

\begin{figure} [t!]
\begin{center}
{\epsfig{file=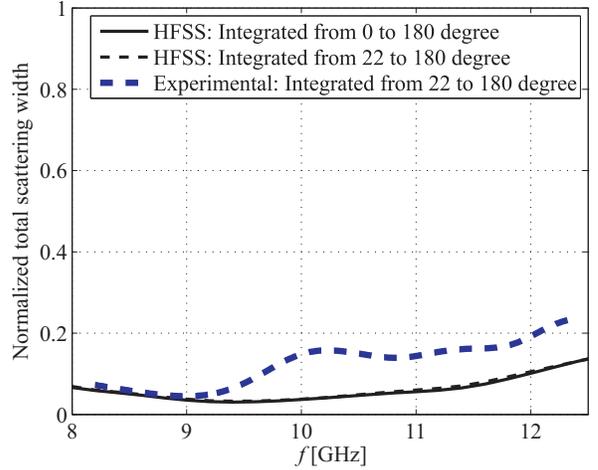, width=0.49\textwidth}}
\caption{Experimental and numerical results for the normalized
total scattering width at X-band.} \label{TSW_exp}
\end{center}
\end{figure}

As explained in the previous section, the measurement setup is
limited in the angular range. To take this limitation into
account, in the following we compare the experimental results to
two sets of numerical results: the one shown in Fig.~\ref{TSW} and
one where the integration in (2) is done from $22^{\circ}$ to
$180^{\circ}$, as in the experiments. The two numerical results
are shown together with the experimental result in
Fig.~\ref{TSW_exp}. We can conclude that the two numerical results
agree very well. This is explained by the fact that for both
uncloaked and cloaked objects most of the scattering occurs in the
forward direction (close to $\phi=180^{\circ}$) and scattering at
low angles insignificantly contributes to the normalized total
scattering width. Moreover, the scattered field amplitude for both
the uncloaked and cloaked objects is almost constant with respect
to $\phi$ for $\phi=0^{\circ}...22^{\circ}$.

The experimental results shown in Fig.~\ref{TSW_exp} demonstrate
that the realized cloak works effectively in the whole X-band, but
the cloaking effect is not quite as strong as in the numerical
results. However, the broadband cloaking effect is clearly
confirmed since the numerical results indicate that the cloak
reduces the total scattering width by more than 90~\% within the
frequency band from 8.2~GHz to 11.9~GHz and the experimental
results show reduction of more than 80~\% in the same band.

Part of the disagreement between the numerical and experimental
results can be due to measurement and fabrication inaccuracies and
the non-ideal orientation of the cloak structure. Each of the
transmission-line grids (each grid comprising one half of the
parallel-strip transmission line) are manually inserted on top of
each other and due to the insulation material between the adjacent
grids, it is difficult to align the transmission-line strips
exactly. We found that a positioning error of a few degrees can be
easily introduced to the alignment and this was found by numerical
simulations to result in deterioration of the cloaking effect in
an amount comparable to the results shown in Fig.~\ref{TSW_exp}.

\begin{figure} [t!]
\begin{center}
{\epsfig{file=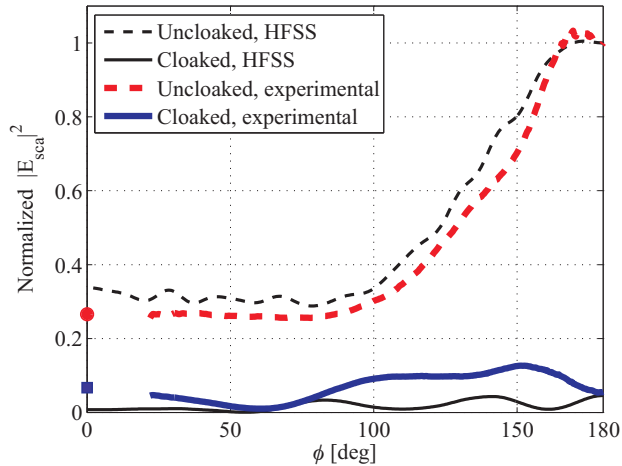, width=0.49\textwidth}}
\caption{Experimental and numerical results for the intensities of
scattered fields as functions of $\phi$, at the frequency 10~GHz.
The intensities are normalized to the uncloaked case at
$\phi=180^{\circ}$. The circle and square represent the results of
the monostatic measurements for uncloaked and cloaked objects,
respectively.} \label{10GHz}
\end{center}
\end{figure}

In addition to the normalized total scattering width it is
interesting to take a look at the angular dependency of scattered
fields in the cases of cloaked and uncloaked objects
(Fig.~\ref{10GHz}). This analysis reveals the directions into
which the objects scatter most. Here we fix the frequency to
10~GHz, i.e., close to the optimal cloaking frequency, and plot
the scattered field intensities obtained from (1) as functions of
the angle $\phi$. To illustrate the cloaking effect, we normalize
all the scattered field intensities to the intensity of the
uncloaked object at $\phi=180^{\circ}$. As explained in Section
III, we can also use the monostatic measurement and (3) to confirm
the levels of backscattering in uncloaked and cloaked cases.
Figure~\ref{10GHz} presents these results. Compared to the
numerical results, the experimental curve in the cloaked case
shows somewhat higher levels of scattering close to the forward
direction, which evidently leads to higher level of the normalized
total scattering width shown in Fig.~\ref{TSW_exp}. The monostatic
measurements (marked with square and circle in Fig.~\ref{10GHz})
confirm that the scattering is quite stable with respect to the
angle in the range from $0^{\circ}$ to $22^{\circ}$.

\section{Conclusions}

Bistatic measurements have been conducted in order to
experimentally characterize the reduction of scattering from a
metal object with a volumetric transmission-line structure used as
an electromagnetic cloak at X-band. The characterization is done
by comparing the measured total scattering widths of the object
alone to the case when the same object is inserted inside the
cloak. The results demonstrate a wide cloaking bandwidth. The
total scattering width of the metal object is reduced by more than
80~\% in a relative bandwidth of 40~\% (8~GHz -- 12~GHz). The
measured results are compared to numerical results and these are
in fairly good agreement with each other, although the numerical
results indicate even slightly better cloaking performance.

\end{document}